\documentstyle[12pt]{article}
\input epsf.tex
\title{Dynamical Scaling Behavior of the Swift-Hohenberg Equation Following a
Quench to the Modulated State}
\author{Qing Hou, Shin-ichi Sasa, Nigel Goldenfeld\\
University of Illinois, Urbana, IL 61801}
\date{}
\begin{document}
\maketitle

\begin{abstract}

We study the kinetics of phase transitions in a Rayleigh-Benard system after
onset of convection using 2D Swift-Hohenberg equation. An initially uniform
state evolves to one whose ground state is spatially periodic. We confirmed
previous results which showed that dynamical scaling occurs at medium quench
($\epsilon = 0.25$) with scaling exponents $1/5$ and $1/4$ under zero noise and
finite noise respectively. We find logarithmic scaling behavior for a deep quench
($\epsilon = 0.75$) at zero noise. A simple method is devised to measure the
proxy of domain wall length. We find that the energy and domain wall length exhibit
scaling behavior with the same exponent. For $\epsilon = 0.25$, the scaling
exponents are $1/4$ and $0.3$ at zero and finite noise respectively. 

\end{abstract}

\hspace{0.3cm}
Classification: PACS:47.54.+r;47.27.Te

\hspace{0.3cm}
Keywords: scaling, pattern formation.

\vspace{0.4cm}
\section{Introduction}

The dynamics of pattern formation in systems away from equilibrium is a
fascinating subject to study. There are numerous examples that exist in nature,
such as Rayleigh-Benard convection, chemical reaction and biological pattern
formation etc.  Much work \cite{crossone} has been done on Rayleigh-Benard
convection due to its relative simplicity. There, a fluid is confined between
two horizontal plates which are heated from below. When the temperature
difference is big enough such that the Rayleigh number $R$ exceeds a critical
value $R_{c}$, the fluid in the uniform state becomes unstable to one which
consists of spatially periodic convection rolls. These rolls form domains with
a typical size which grows with time. Swift and Hohenberg developed a simple
model \cite{swift} to describe this process. The order parameter equation
reads,

$$\frac{\partial \psi}{\partial t} 
= \epsilon \psi - \psi^{3} - (1 + {\it\Delta})^{2}\psi + \eta$$

\noindent  
where $\psi$ is related to the vertically averaged magnitude of the velocity
component normal to the plate. The reduced Rayleigh number $\epsilon = (R -
R_{c})/R_{c}$ measures how far the system is above the onset of convection.
The thermal noise $\eta$ satisfies the usual relation $\langle \eta(x,t)
\eta(x',t') \rangle = F \delta(x - x') \delta(t - t')$. This model captures the
three basic features of pattern formation in the Rayleigh-Benard system,
namely, initial growth of the instability, its nonlinear saturation and
development of a spatially periodic pattern.

Away from the threshold where $\epsilon \sim O(1)$, one can perturb around the
ground state $\psi_{0}(x)$ which is periodic in $x$ and define a phase $\phi$ as
in

$$\psi = \psi_{0}[\phi({\bf x}, t)] + O(\zeta),$$
 
\noindent
where $ {\bf \nabla} \phi({\bf x}, t) = {\bf q}({\bf x}, t)$ and the gradients
of local wavenumber ${\bf q}$ are of the order of a small number $\zeta$. Thus
one obtains the {\it phase equation} \cite{pomeau,passot}. To lowest order in
$\epsilon$,

$$ \partial_{t} \phi =  D_{\perp}\partial^{2}_{\perp}\phi + 
D_{\|}\partial^{2}_{\|}\phi. $$

\noindent 
where $\perp$ and $\|$ denotes directions along and normal to the rolls and the
$D$'s are diffusion coefficients. Direct dimensional analysis tells us that the
characteristic length grows as $t^{1/2}$. However, $D_{\perp} \rightarrow 0$
due to local wavenumber relaxation. The next order expansion gives us a fourth
order gradient term and one expects  a scaling behavior $L \sim t^{1/4}$ in the
transverse direction (along the rolls) \cite{crossone,cross}  which physically
corresponds to curvature relaxation \cite{elder}.

Scaling behavior is seen in numerical simulations. However, the exponent is
$1/5$ for noiseless case \cite{elder,crosstwo} and is apparently changed to $1/4$ when
finite noise is added \cite{elder}. Non-potential variants of the
Swift-Hohenberg system which take into account of mean field flow give the same
result \cite{crosstwo}  (non-uniform horizontal motion of the fluid, important
for fluid with small viscosity).

The fact that the phase equation predicts a different value of exponent
from simulation and that the exponent is noise dependent is intriguing.
Actually, since the phase equation starts from perturbation around an
ideal state, it fails to encompass the many defects that are present in
a real system. However, the coarsening process in pattern formation is
brought about by defect movement. Through defect annihilation, larger
domains of nearly ideal configuration form at the expense of smaller
ones. This important role played by defects is well-known; a good
example is the kinetics of the formation of the Meissner state in a type
II superconductor at zero field \cite{liu}, where the inter-vortex
potential $U(r)$ determines the scaling behavior of the inter-vortex
distance.

\section{Numerical Results}

We approach the problem primarily as numerical experimentalists. We try to
measure strategic quantities which could shed light on the mechanism of the
pattern coarsening, especially quantities which could be related to defects in
the system. We use an explicit, first order accurate, pseudospectral method.
The boundary is periodic to simulate large aspect ratio systems. Two $\epsilon$
values are used. For $\epsilon = 0.25$, the lattice is 512 by 512, time step
size if 0.1. For $\epsilon = 0.75$, the lattice is 256 by 256, time step size
is 0.05. In either case,  There are 8 lattice points per ideal period. In the
case of $\epsilon = 0.25$, we also did a case with finite noise to study the
effect of noise on the dynamics. To obtain the statistics, each run is repeated
with 10 different initial random configurations to average over.

\vspace{0.3cm}
\noindent
{\bf A. $\epsilon = 0.25$. }

We measure three quantities. One is the structure factor $S(k)$. This is the
canonical measure from which one can calculate the dynamical scaling exponents.
We also measure the energy of the system as well as the sum of all the domain
wall lengths.  The former could signal, in this relaxational model, whether
bulk or defects are contributing most to energy relaxation, while the latter
could give us information on defect dynamics.

The structure factor is defined as the statistical average of the fourier
spectrum of the equal time correlation function, averaged over all directions
of the 2D wave vector,

$$S(k) = {\it FT}\cdot\langle \psi({\bf x},t) \psi({\bf x}',t') \rangle. $$

\noindent
The Swift-Hohenberg system belongs to the I$_{s}$ instability class
\cite{crossone}. Accordingly, $S(k)$ is peaked at $k = k_{0}$ away from the origin, corresponding to the ground state and most unstable mode above onset of
convection. Due to the nonlinear interaction, there are peaks corresponding to
higher harmonics present in $S(k)$. Away from these higher harmonics, it is
observed that $S(k)$ is a function of $|k^{2} - k^{2}_{0}|$. We fit $S(k)$ to
squared Lorentzian form \cite{crosstwo}

$$ S(k) = \frac{S(k_{0})}{(1 + ({\it\Gamma}(k^{2} - k_{0}^{2})^{2})^{2})}$$

\noindent
where $S(k_{0})$ is the peak amplitude and ${\it\Gamma}$ is a fitting parameter
from which one can calculate $L$, the characteristic width of $S(k)$. The two
parameters $S(k_{0})$ and $L$ exhibit scaling behavior. The exponents are
consistent with results obtained by earlier workers.

\begin{eqnarray}
L \sim S(k_{0}) \sim t^{\phi_{L}}, \mbox{ where } \phi_{L} = 
\left\{\begin{array}{ll}
1/5, & F_{d} = 0, \\
1/4, & F_{d} = 0.05.
\end{array}\right.\end{eqnarray}

\noindent
where we define the discretized noise $F_{d} = F/(\Delta x)^{2}$, where $\Delta
x = \pi/4$ is mesh size used in our simulation. The data plots for $S(k_{0})$
and $L$ are shown in Fig 1.

\vspace{-2.7cm}
\centerline{\epsfxsize=7cm \epsfbox{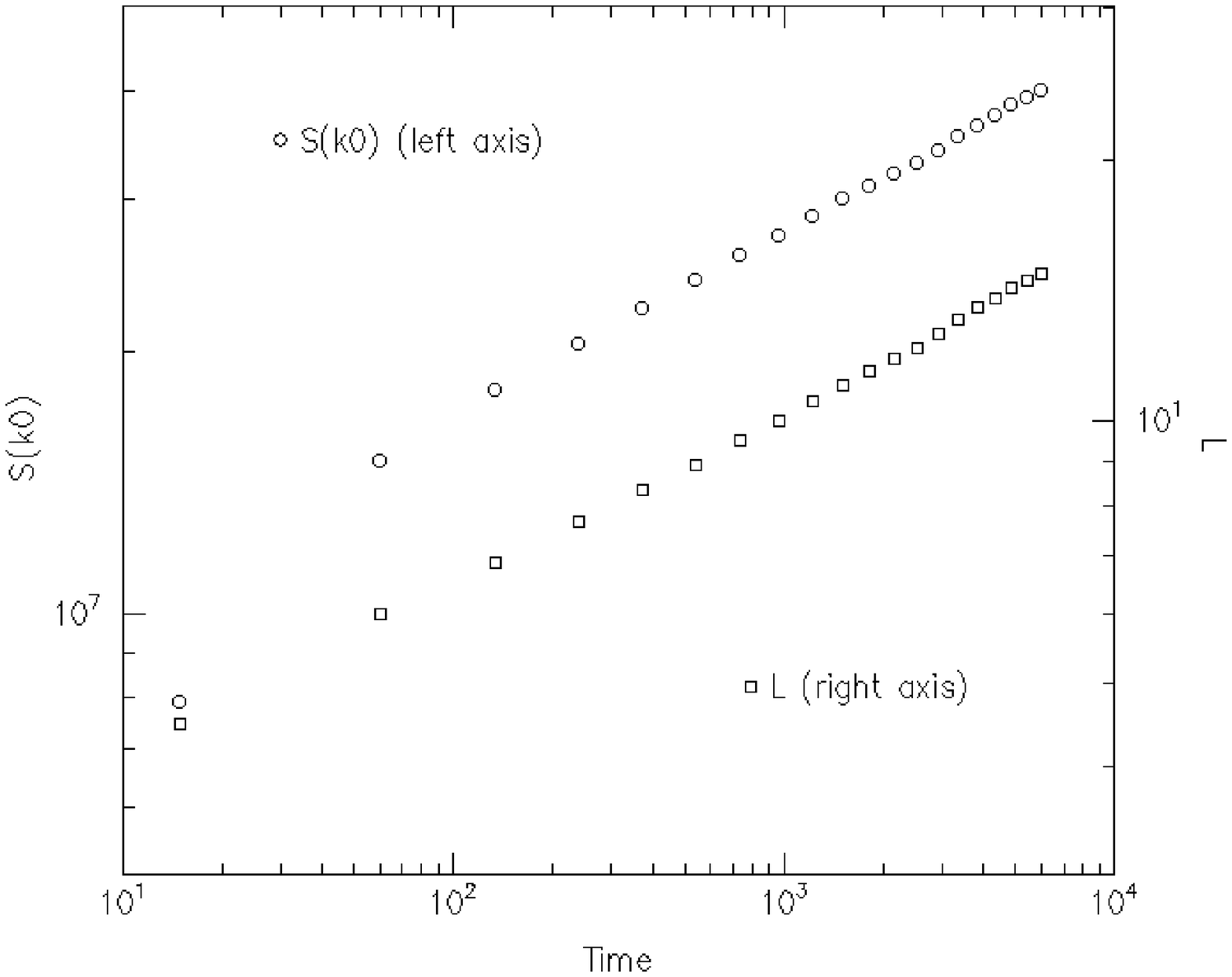} \hskip 0.4cm
            \epsfxsize=7cm \epsfbox{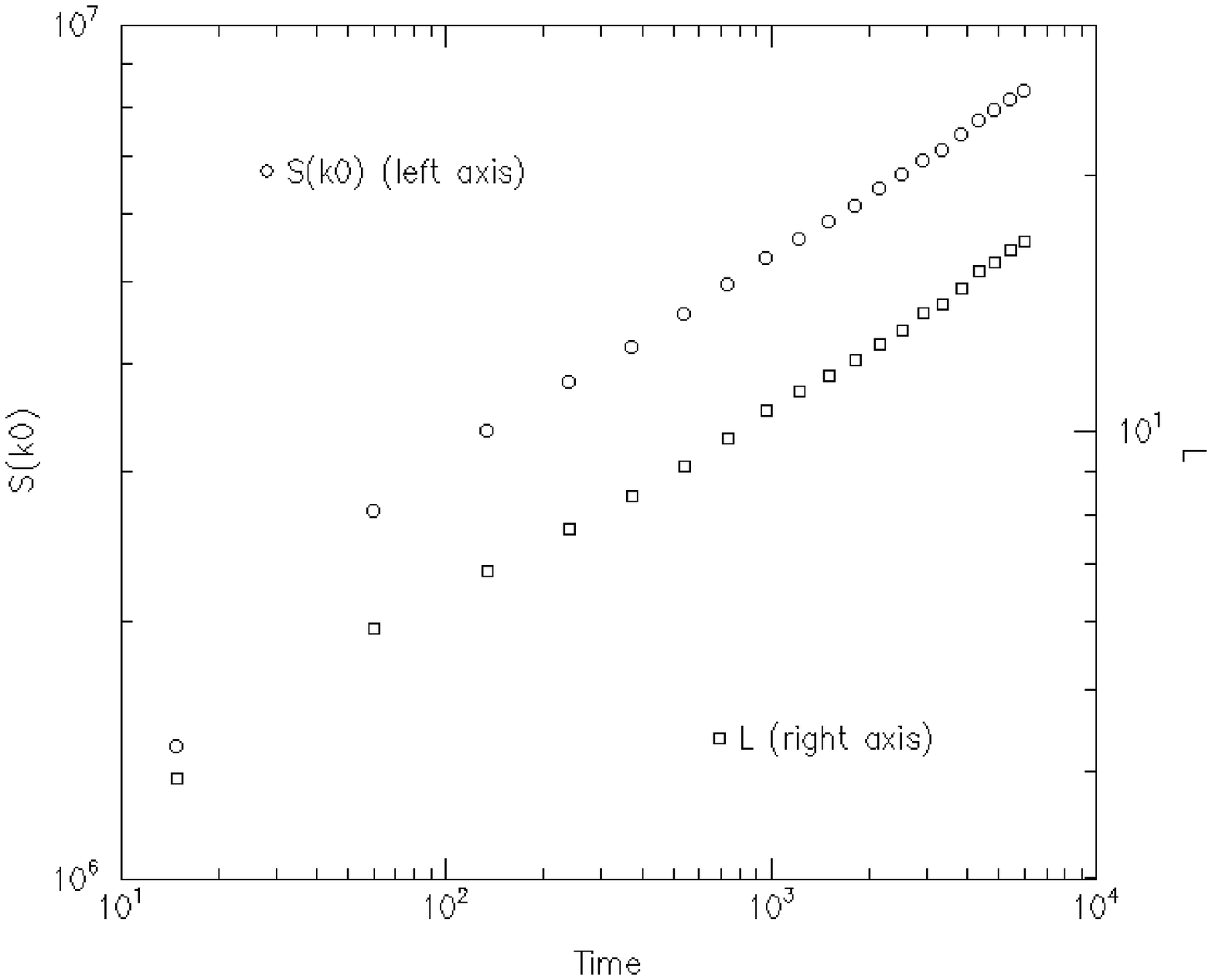}}

\vspace{1.5cm}
\hspace{0.4cm}{\small Fig 1. L/S(k$_{0}$) vs Time ( F$_{d}$ = 0 left and 
F$_{d}$ = 0.05 right). Left and right reference

\hspace{0.4cm}lines have slopes 1/5 and 1/4 respectively. }

\vspace{.3cm}
Using these characteristics, we scale the structure factor from different times.
The central region of $S(k)$ collapses onto a universal curve (Fig 2).

\vspace{-3.cm}
\centerline{\epsfxsize=7.5cm \epsfbox{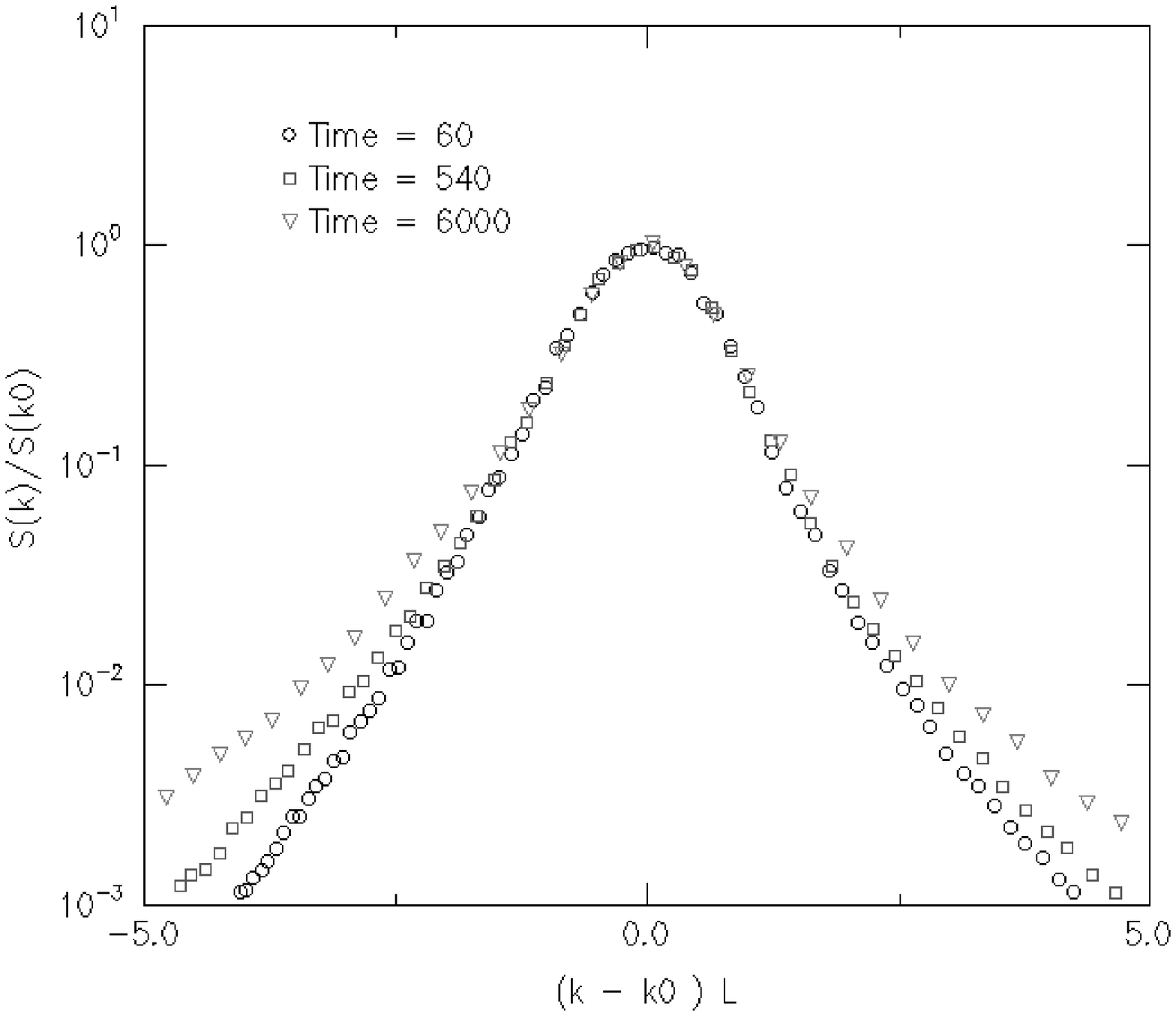}}

\vspace{1.5cm}
\hspace{3.5cm}{\small Fig 2. Scaled S(k) vs k$^{2}$ - k$_{0}^{2}$. Noiseless.} 

\vspace{0.3cm}
\noindent 
It is interesting to notice that $S(k_{0})$ scales as $L$ \cite{elder}
not $L^{2}$ as is the usual case for phase ordering in a 2D system. This
is due to the fact that, in the process of Fourier transformation
$S(\vec{r}) {\rightarrow} S(\vec{k})$, most of the contribution comes
from annulus $k_{0} dk$ instead of area $k dk$ at the origin where $k
\sim  dk$. Since ${\it\Gamma}$ is basically $L$, we write the scaling
form of $S(k)$ as

$$S(k) = L l \cdot f(L l |k^{2} - k^{2}_{0}|) $$

\noindent
where $l$ is a constant length to give the right dimension, proportional to
the periodicity.

There is an energy functional $H$ in our model, where 

$$ H = \frac{1}{2}[(1 +
{\it\Delta})\psi]^{2} - \frac{\epsilon}{2}\psi^{2} + \frac{1}{4}\psi^{4}$$

\noindent
The system evolves in a way such that the energy is decreasing as a function of
time. The questions then are: is there energy scaling? If so, what is the
scaling exponent? And is it related to the scaling of the characteristic length
$L$? Scaling in energy was observed in other systems such as phase ordering in
nematic films \cite{zap} where it is found to be mostly due to bulk energy
relaxation.

The ground state energy density can be easily calculated to be $-\frac{1}{6} 
\epsilon^{2}$ to the lowest order in $\epsilon$. In the noiseless case, this is
also the long time limit. We can measure the excess energy $E$ of the system
relative to that of its ground state. For the noisy case, it is a bit tricky,
because there is a \lq thermal' energy component from the external noise
source. So the long time limit is unknown. What we did was to determine this
value empirically by fitting the energy evolution to a power law allowing for
an offset. Then excess energy is found by subtracting this offset. Whether this
is a good fit is answered by seeing whether the scaling prevails over several
decades.

Indeed, a nice scaling regime is found. However, the scaling exponent is not
related to that of the characteristic length $L$ in any simple way (such as
$L^{-2}$ as expected from ordinary bulk relaxation). We think the dominant
contribution comes from defects as will be discussed below. Our results are
shown in Fig 3, and may be summarized as:

\begin{eqnarray}
E \sim t^{-\phi_{E}}, \hbox{ where } \phi_{E} = 
\left\{\begin{array}{ll}
1/4 &, \mbox{ } F_{d} = 0, \\
0.3 &, \mbox{ } F_{d} = 0.05.
\end{array}\right.\end{eqnarray}

\vspace{-3cm}
\centerline{\epsfxsize=7cm \epsfbox{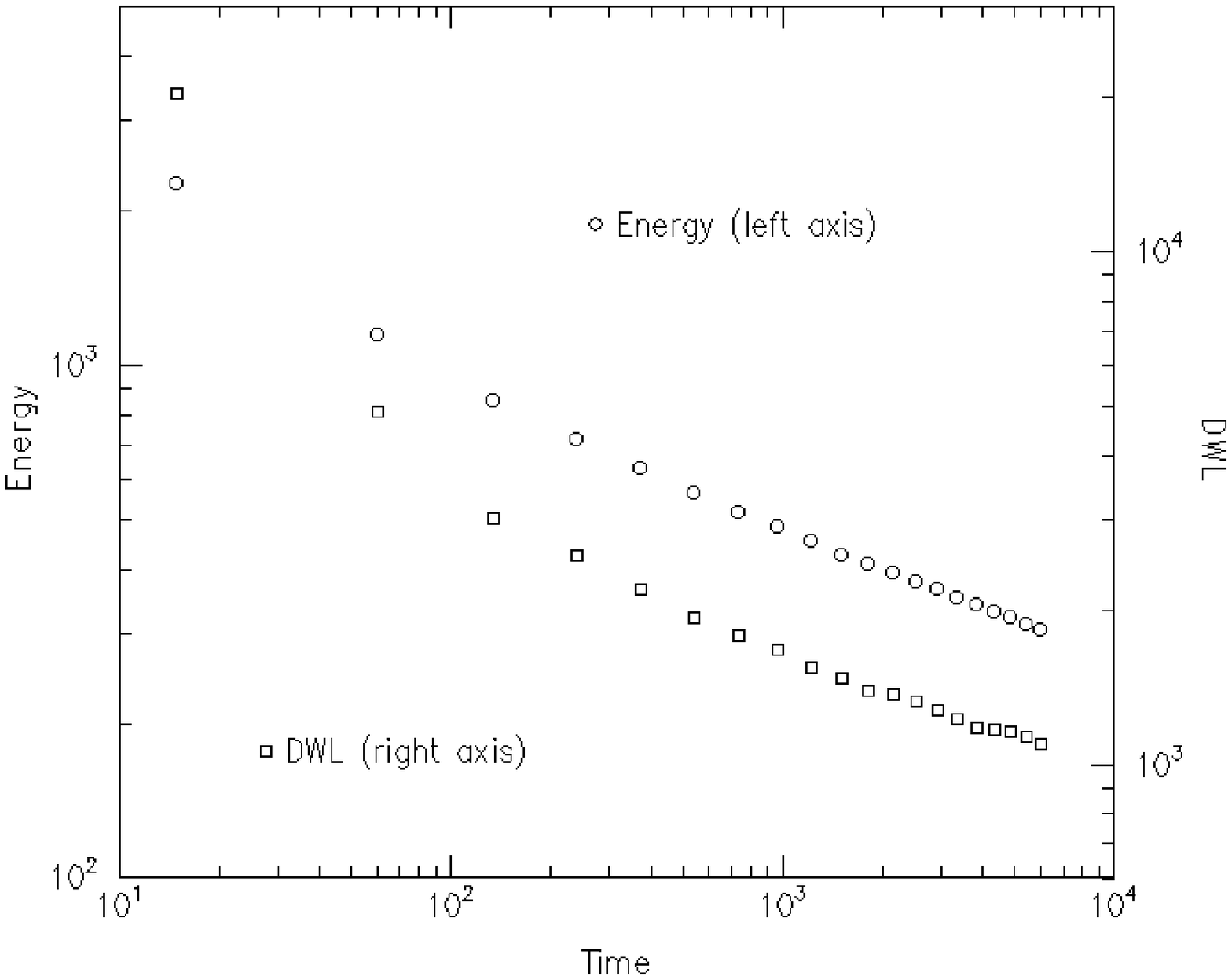}  \hskip 0.4cm
            \epsfxsize=7cm \epsfbox{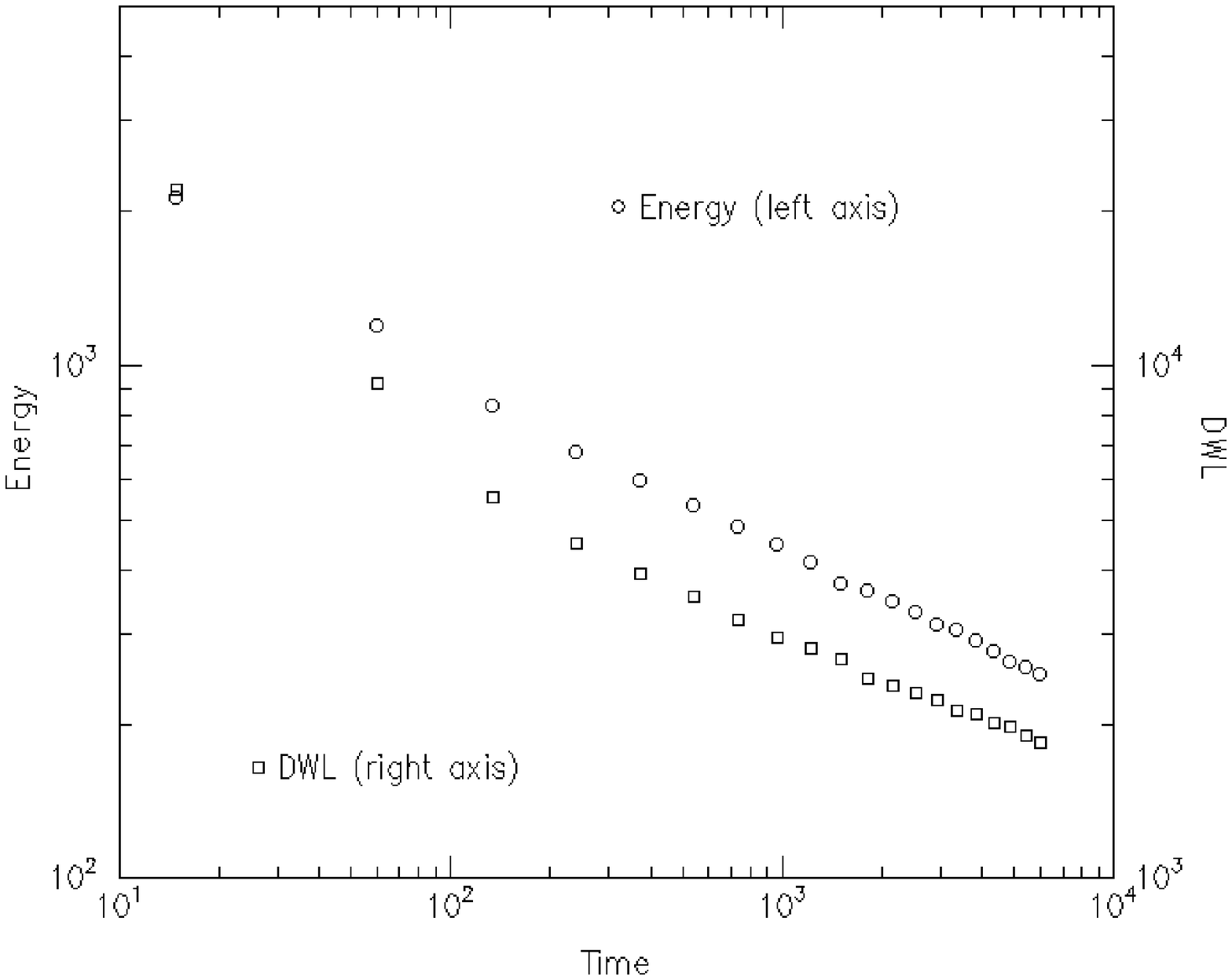}}

\vspace{1.5cm}
\hspace{0.4cm}{\small Fig 3. E/DWL vs Time (F$_{d}$ = 0 left and F$_{d}$ = 0.05
right).  Left and right reference

\hspace{0.4cm}lines have slopes 1/4 and 0.3
respectively.}
  
\vspace{0.2cm}
An accurate measurement of domain wall lengths should be very helpful in
clarifying one of our main concerns: what is the role of defects? However, it
is quite difficult to determine where a domain wall is located mathematically.
Other researchers have used optical filtering procedures to visualize domain
walls \cite{crossfour}. However, no attempt has been made to measure the length
of these domain walls. Furthermore, the filtering procedure is very computationally
intensive. 

We devised a simple method to identify domain walls and obtained a proxy of the
domain wall lengths. The method is based on the observation that, within a
domain, the order parameter configuration is close to the ground state
solution, $\psi \sim A \sin({\bf k_{0}}\cdot{\bf x})$. We have ${\bf \nabla}\psi \sim A {\bf k_{0}} \cos({\bf k_{0}}\cdot{\bf x})$ and   $A^{2} \sim
\psi^{2} + ({\bf \nabla}\psi)^{2}/k_{0}^{2}$. The converted amplitude $A$ is almost constant everywhere (Fig 4) except at grain boundaries, which can thus be

\vskip 0.2cm
\centerline{\epsfxsize 5.2cm \epsfbox{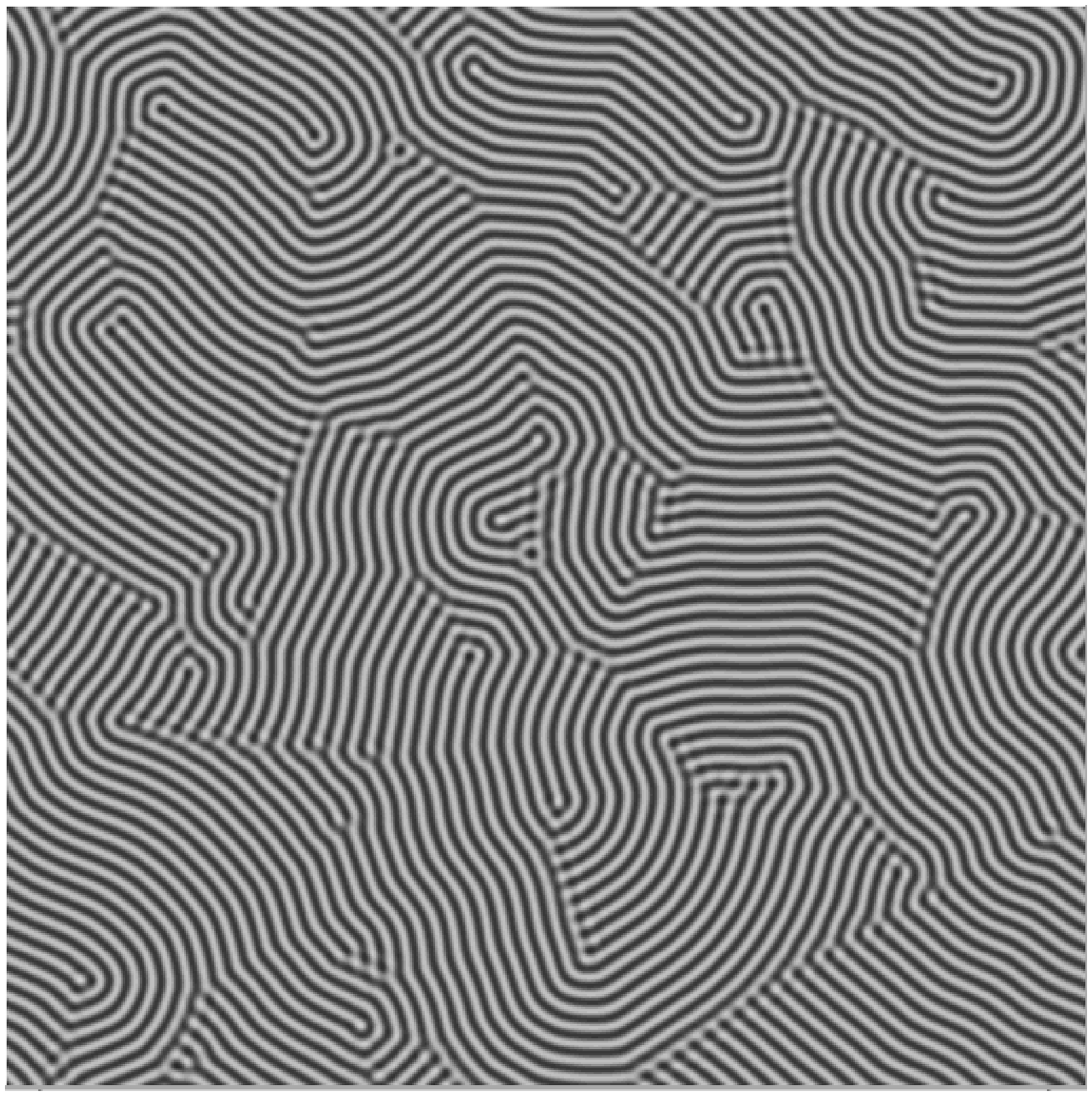}  \hskip 1.5cm
            \epsfxsize 5.2cm \epsfbox{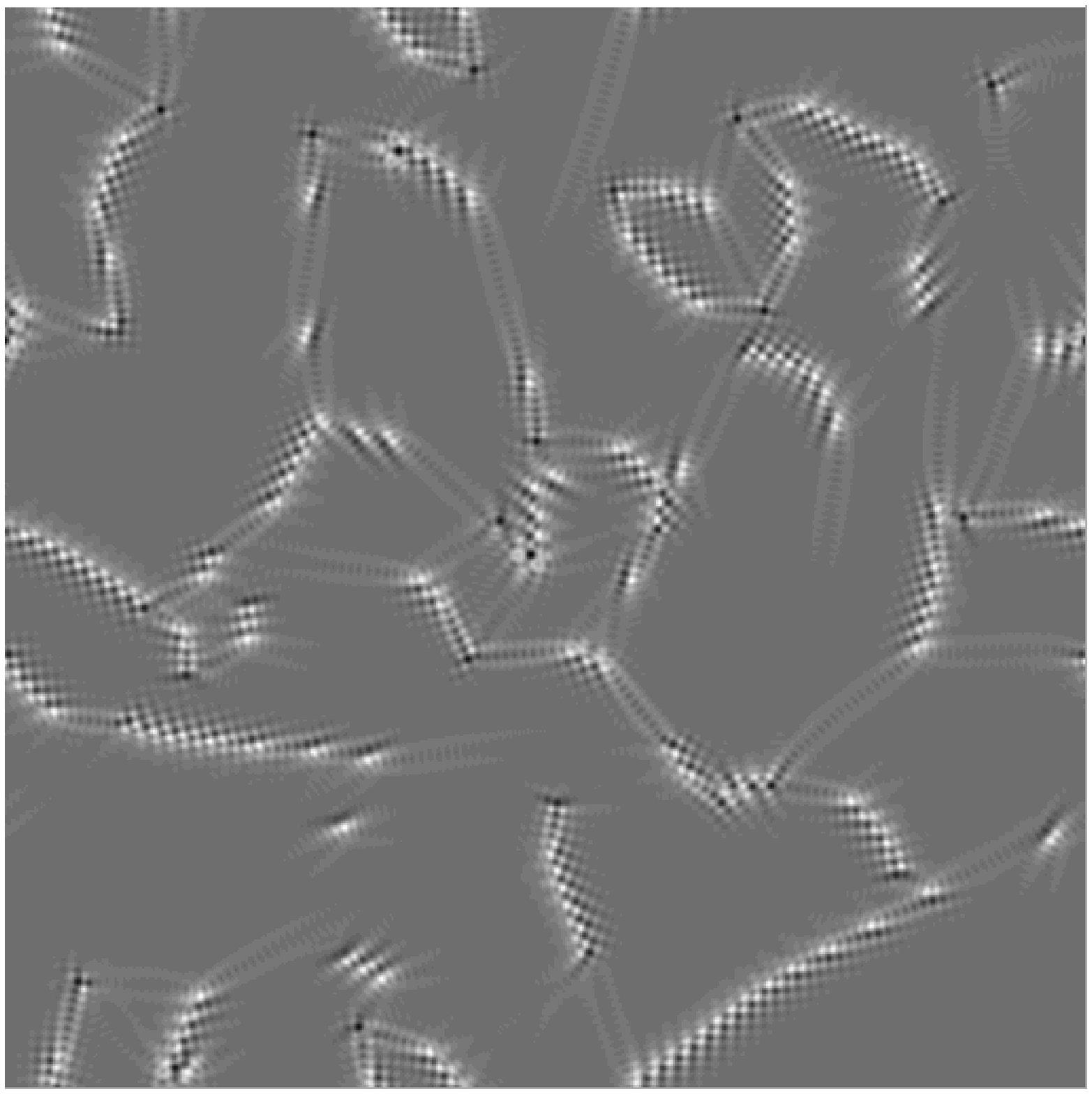}}

\hspace{0.4cm}{\small Fig 4. Order parameter configuration (left) and
converted amplitude A squared (right).    }

located by filtering with respect to $A$. We set up a threshold so that, if the calculated $A^{2}$ is bigger than $0.7 \times Max(A^{2}) + 0.3 \times AVG(A^{2})$ or smaller than $0.7 \times Min(A^{2}) + 0.3 \times AVG(A^{2})$, that point is counted as belonging to a domain wall. The length proxy $DWL$ is defined as a count of all such points.

Scaling behavior is observed for $DWL$. The exponent is different from that of
the characteristic length $L$. Instead, it is the same as that of the energy
relaxation (Fig 3)($DWL$ for noiseless case was an average of 5 runs since we
devised the above way of measuring $DWL$ after the first 5 runs).
This strongly suggests that excess energy is primarily distributed at domain walls and other defects. This part of the energy relaxes
slowly, via defect annihilation. On the other hand, bulk energy is the fast
mode and scales as $L^{-2} \sim t^{-2/5}$ and decays much faster than the
excess energy $E$. At this point, it is tempting to conclude that defects are 
indeed the driving force behind the coarsening process due to its dominant
contribution to the excess energy. However, it is still not clear how one could
link the evolution of $E$ and $DWL$ with the growth of $L$. 

\begin{eqnarray}
DWL \sim t^{-\phi_{DWL}}, \mbox{ where }\phi_{DWL} = 
\left\{\begin{array}{ll}
1/4&, \mbox{ } F_{d} = 0, \\
0.3&, \mbox{ } F_{d} = 0.05.
\end{array}\right.\end{eqnarray}

\vspace{0.5cm}
\noindent
{\bf B. $\epsilon = 0.75.$}

For large $\epsilon$, the system is much further away from equilibrium. We
anticipated that there would be more isolated dislocations and they could give
rise to a different coarsening process. This is indeed what we found.  The
system evolves slowly. The characteristic length exhibits logarithmic scaling
behavior, while the excess energy decays even slower and the number of defects
stays almost constant (Fig 5). 

$$ L \sim S(k_{0}) \sim \log{t}, \hbox{ } F_{d} = 0.$$

\vspace{-2.7cm}
\centerline{\epsfxsize 7truecm \epsfbox{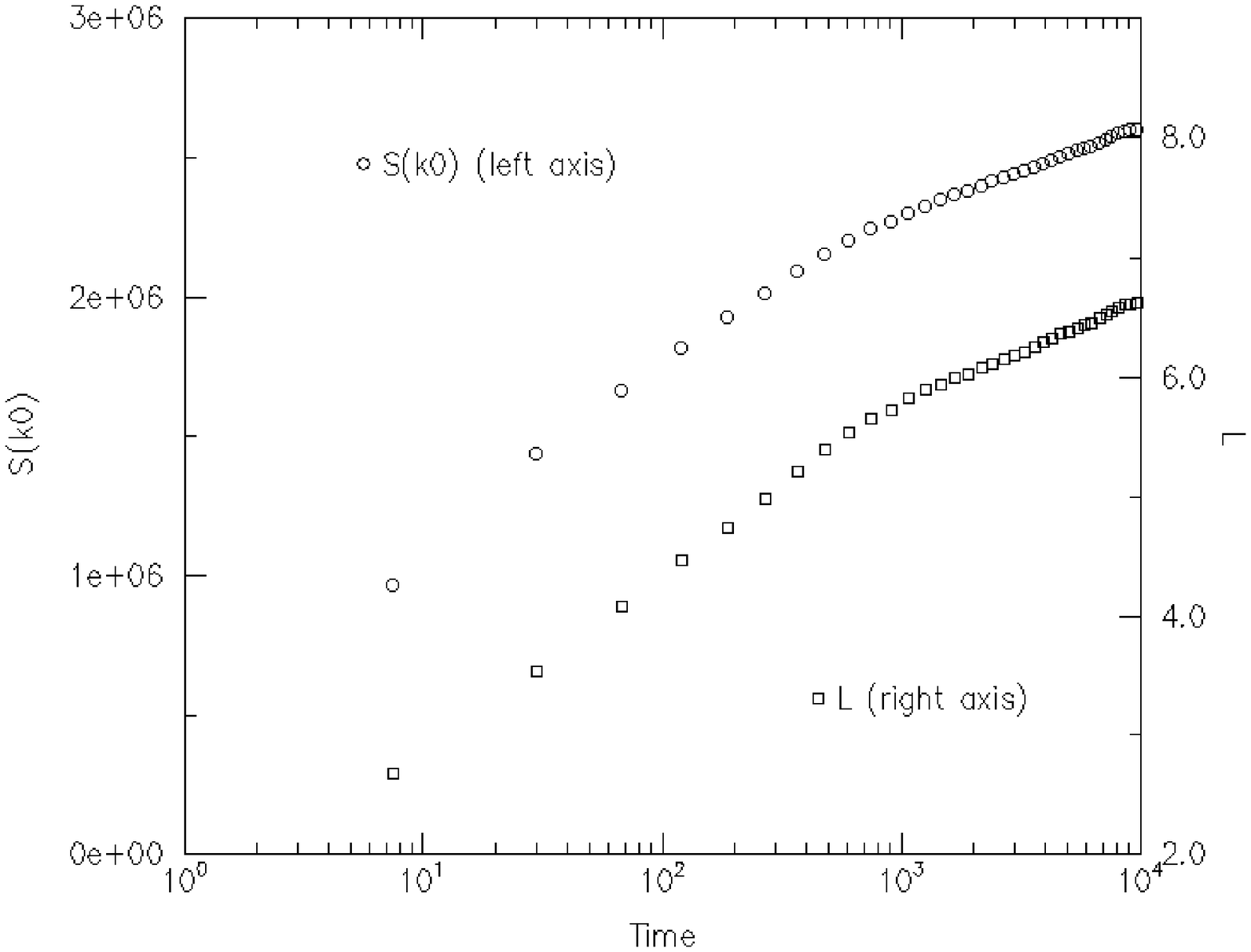}\hskip 0.4cm
            \epsfxsize 7truecm \epsfbox{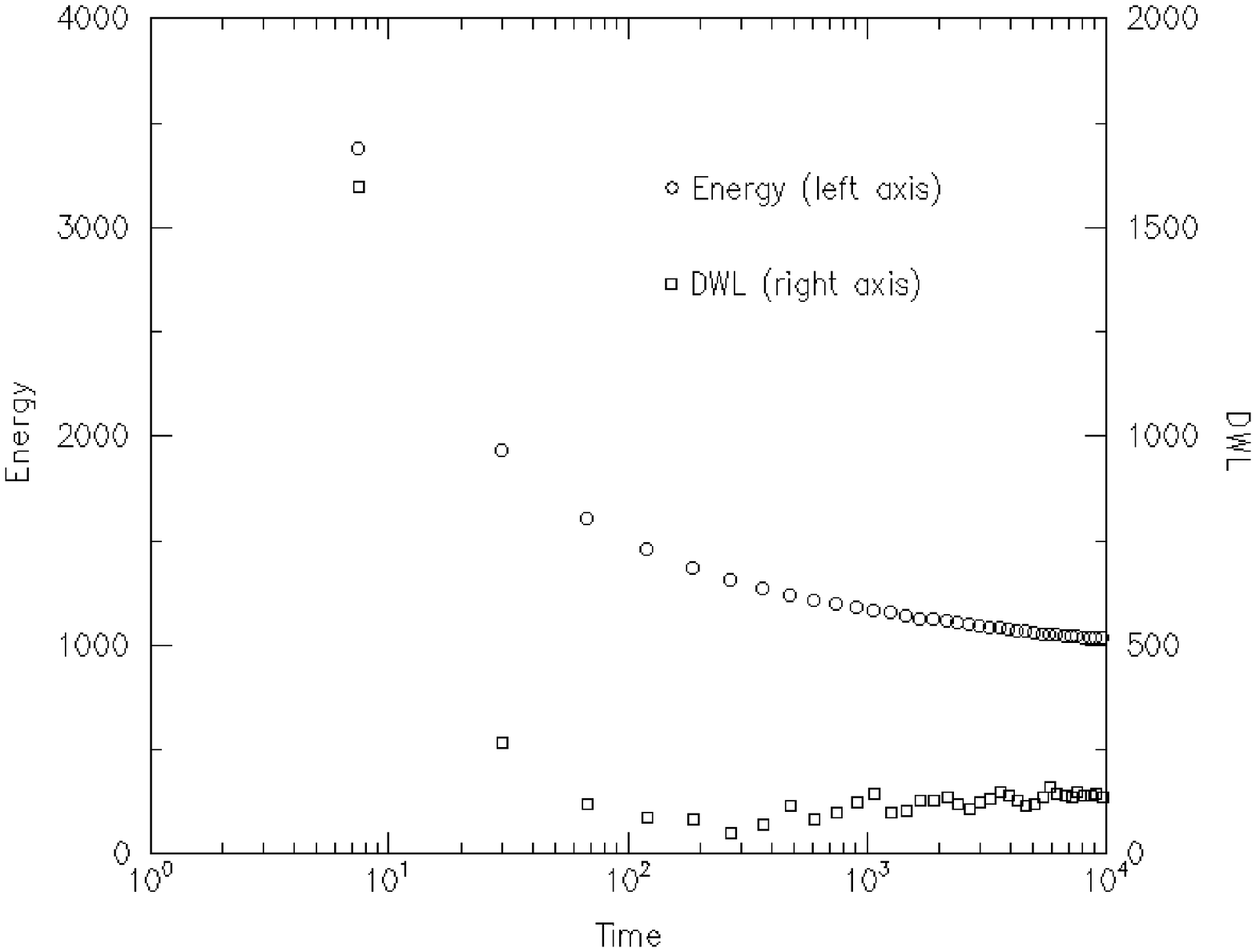}}

\vspace{1.5cm}
\hspace{0.4cm}{\small Fig 5. L/S(k$_{0}$) vs Time (left) and
E/DWL vs Time (right)}

\vspace{0.2cm}
A heuristic argument for the logarithmic behavior uses the fact that the pair
potential \cite{toner} between two dislocations is exponentially decaying with
respect to the distance between them, namely,

\begin{equation}
U = \frac{U_{0}}{\sqrt{z/\lambda}} \exp(-\frac{x^{2}}{z \lambda})
\end{equation}

\noindent
Setting $r \sim x \sim z$, one has, for the damped defect motion, 

\begin{equation}
\eta \frac{d r}{d t} = -\frac{d H}{d r} = -\frac{v_{0}}{\sqrt{r/\lambda}} 
\exp(-\frac{r}{\lambda}).
\end{equation}

\noindent
The leading order of the solution to above equation gives us logarithmic 
behavior. Now substitute this leading order behavior $r \sim \log(t)$ into the
defect pair potential energy, one finds that the pair potential energy part
scales as $t^{-1}$, much faster than what one sees in the $E$ vs $Time$ plot
shown above. Thus, there must be other significant energy sources which could
be some energy background not important to dynamics.

\section{Conclusion}

In this paper, we presented results from our numerical simulation of the
Swift-Hohenberg system. For a medium quench ($\epsilon = 0.25$), the
system has many domain wall defects. We confirmed the dynamic scaling
behavior in structure factor observed by previous researchers. We showed
that the energy of the system also exhibits scaling behavior. Its
scaling exponent does not have a simple relationship with that of the
characteristic width, making it difficult to draw any concrete
conclusion. However, one notices that the energy relaxation is slower
than what one would expect from bulk contribution, namely, $L^{-2}$.  We
devised a simple way to measure the proxy of domain wall length. The
$DWL$ has the same scaling exponents as the energy, suggesting that
energy is concentrated on defects. This is consistent with the notion
that defects are the driving force behind the coarsening process. We
find that adding noise into the system speeds up the evolution, because
all the exponent values increase.

For a deep quench ($\epsilon = 0.75$), we see a much slower coarsening
process where there are mostly isolated dislocations in the system. The
characteristic length scale shows logarithmic behavior. To explain this,
we give a heuristic argument assuming overdamped isotropic dynamics
under the influence of the pair potential between
defects.
                                    
\section*{Acknowledgement}
This work is supported by grant NSF-DMR-93-14938 and NSF-DMR-89-20538.

\end{document}